\documentclass{Rinton-P9x6}

\begin{document}

\title{Towards a heat kernel expansion for the electromagnetic field
interacting with a dielectric body of arbitrary form}

\author{Irina Pirozhenko}

\address{Bogoliubov Lab. of Theoretical Physics, Joint Institute for Nuclear Research,
141980 Dubna, Russia\\
E-mail: pirozhen@thsun1.jinr.ru}


\maketitle

\abstracts{
%
The results on the heat kernel expansion for the electromagnetic field in the
background of dielectric media are briefly reviewed. The common approaches to the
calculation of the heat kernel coefficients are discussed from the viewpoint
of their applicability to the electromagnetic field interacting with dielectric
body of arbitrary form. Using the toy-model of scalar photons we develop
multiple reflection expansion method which seems the most promising one when
the field obeys dielectric-like matching conditions on an arbitrary interface
and  show that the heat kernel coefficients are expressible through geometric
invariants of the latter.}

\section{The review of results and approaches}
Studying the heat kernel $K(x,y;t;L)=<x|\exp(-t\,L)|y> $
provides us with various information concerning operator $L$.
Of special interest for the QFT under the influence of
external conditions  is the heat kernel
expansion in powers of the small parameter $t$
\begin{equation}
\bigl.K(t|L_D)\bigr|_{t\to0} \sim(4\pi
t)^{-D/2}\sum_{n=0}^{\infty}t^{n/2} B_{n/2},
\label{eq:hke}
\end{equation}
as the coefficients $B_{n/2}$  govern the short distance behaviour of the propagator
and define one-loop divergences and counterterms. Moreover the heat kernel is a powerful
tool for analysing quantum anomalies and  various perturbative expansions
of  effective action (see \cite{Vass} and references therein).

To derive the heat kernel expansion for the electromagnetic field interacting with
dielectric body of arbitrary form is an objective motivated primarily by
the necessity of clarifying  the situation with the Casimir effect in
dielectrics. Many authors have devoted their efforts to the Casimir
energy calculations for dielectric bodies, the results appeared to be
controversial.

The electromagnetic field in unbounded media with  dielectric
permittivity $\varepsilon$ and magnetic permeability $\mu$
smoothly depending on  coordinates may be treated as propagating in the curved
space  with the effective metric defined by $c(x)^2=1/({\varepsilon(x)\mu(x)})$.
Then the standard formulae\cite{Gilkey} expressing the local coefficients
$B_{k}(x,x)$ through  the polinomials of manifold's geometric invariants are
valid~\cite{BKV}. As the medium is unbounded the coefficients with
half-integer numbers are absent in this heat kernel expansion.
From the heat kernel expansion obtained in~\cite{BKV} it follows that there is no
cancellation of ultraviolet divergences between ghosts and "nonphysical"
components of the vector potential. Moreover the coefficient $B_{2}$ whose
vanishing in 3-dimensional space could assure the uniqueness of the
renormalisation for one-loop divergences is nonzero.

If the dielectric body possesses a curved boundary, then
the derivation of the heat kernel expansion is getting cumbersome.
By now the expansions were obtained only for the  boundaries with
high symmetry. Dielectric balls and cylinders were considered in~\cite{BKV1,BP}.
The non-vanishing coefficient $B_2$ is the common feature for material balls
and cylinders. It means that in general case it is difficult to fix
the finite part of the ground-state energy and the results of different
calculations may not coincide. The account for the dispersion~\cite{BorKir}
also does not cure the arbitrariness which remains in the finite part of the
ground state energy after the removal of the diverging contributions.

Among the balls and cylinders there are two known exceptions.
Under special choice of the parameters $\varepsilon$ and $\mu$ for the body and
the surrounding medium, namely when $\varepsilon_1 \mu_1 =\varepsilon_2 \mu_2= c^2$,
all leading heat kernel coefficients except $B_{3/2}$ are equal to zero.
The  second exception is the vanishing of $B_2$ in the dilute approximation,
i.e., to the order $(\varepsilon_1 - 1)^2$ for $\varepsilon_1\to 1$ and
$\varepsilon_2=\mu_1=\mu_2=1$.

Being  concerned with the heat kernel expansion for a dielectric body
of arbitrary shape one is unable to use the spherical or cylindrical symmetry
of the boundary for obtaining the eigenfrequency equations in terms of the
Bessel functions.  Therefore the analysis of the poles for the
corresponding spectral zeta function and subsequent derivation of the heat
kernel coefficients~\cite{BEK,BEKL,BEGK} as the residues in these poles
might hardly be accomplished.

The DeWitt iterative procedure was historically the first one used in QFT to obtain
the heat kernel coefficients for curved manifolds. The generalization of this
method for  the manifolds with boundaries makes it too tedious and moreover
requires certain assumptions regarding the general
form of the heat kernel expansion. When considering the electromagnetic
field with  dielectric matching conditions we are lacking intuition which
guided the authors of ~\cite{McO} in Dirichlet and Neumann case.

In the last decade Gilkey's method, which involves constructing the
heat kernel coefficients of all geometric invariants allowed from dimensional
viewpoint and finding numerical coefficients in front of them, proved to be
the most powerful for manifolds with boundaries.
The advantages of this method emerge as it is applied to problems with
singularities concentrated on the boundary such as $\delta_{\Sigma}$-shaped
background potential (domain walls) or non-smooth normal derivatives of the
metric (brane-world scenario).

Gilkey's technique admits non-smooth derivatives of the metric requiring
however the metric itself  to be smooth. Therefore the method is inapplicable for compound
dielectrics  where the effective metric expressed
trough $\varepsilon(x)$  jumps on the boundary making the problem ill-defined.
One has to replace it by a pair of spectral problems on the sides $M^{\pm}$
of the boundary $\Sigma$ supplied with suitable matching conditions.
After that it seems  reasonable to restate obtained problems in the language
of integral equations and to obtain the heat kernel as a sum of generalized multiple
reflection expansion.

\section{Multiple reflection expansion}
The multiple reflection expansion method for the heat kernel has its origin in
the potential theory for parabolic operators~\cite{Tikhonov}.

The solution of the heat equation with initial condition
$K(x,y;0)=\delta(x,y)$ may be  written in the form
\begin{eqnarray}
K(x,y;t)&=&K^{0}(x,y;t)+ \alpha_1 V(x,y;t)+\alpha_2 W(x,y;t),
\label{eq:poten}\\
V(x,y;t)&=&a^2\int\limits_{0}^{t}d\tau
\int\limits_{\Sigma}^{} dz \ K^{0}(x,z;t-\tau)\mu(z,y;\tau), \nonumber\\
W(x,y;t)&=a^2&\int\limits_{0}^{t}d\tau
\int\limits_{\Sigma}^{} dz \ K^{0}(x,z;t-\tau)
\frac{\overleftarrow{\partial}}{\partial n_{z}}\nu(z,y;\tau) \nonumber
\end{eqnarray}
where $K^{0}$ is the free heat kernel, $V$  and $W$ are called respectively simple
and double layer heat potentials. For Dirichlet (Neunann) problem it is convenient
to choose $\alpha_1=0$ ($\alpha_2=0$).

The double layer potential and normal derivative of the simple layer
potential are discontinuous on the interface $\Sigma$
\begin{eqnarray}
\frac{\partial}{\partial n_x}V(x,y;t)\Biggr|_{\Sigma_{\pm}}&=&
\frac{\partial}{\partial n_x}V(x,y;t)\Biggl|_{\Sigma}\,\mp\,\frac{1}{2}\mu({\bf x},y;t),
\label{eq:poten1}\\
W(x,y;t)\Biggr|_{\Sigma_{\pm}}&=&
W(x,y;t)\Biggl|_{\Sigma}\,\pm\,\frac{1}{2}\nu({\bf x},y;t).
\label{eq:poten2}
\end{eqnarray}
Substituting the solution (\ref{eq:poten}) into the boundary condition under
consideration with account of (\ref{eq:poten1}) or (\ref{eq:poten2}) one
arrives at the integral equation defining the density $\mu$ (or $\nu$) of simple
(or double) layer potential. The iterative solution of this integral equation is
called multiple reflection expansion (MRE), where each $n$-th term corresponds to
the $n$-th interaction with the boundary.

The MRE is convergent for  most of physically reasonable boundary conditions
sometimes despite of the absence of a small
expansion parameter. There may exist various MREs, however as the heat kernel is
unique the summation of all possible true ones should result in the same answer.

One can use the MRE  as a tool for obtaining the heat kernel coefficients on condition
that just a finite number of leading interactions with the boundary contributes to
a coefficient with a finite index.

\section{Dielectric-like toy model: scalar photons}
Here we  construct the MRE for the heat kernel of massless scalar field propagating
in conformaly flat 3D-space with conformal factor behaving like a step-function
as some interface $\Sigma$ is crossed (scalar photons in compound dielectric).
The heat kernel $K(x,y;t)$ may be decomposed in four parts
depending on the position of the points $x$ and $y$
\begin{eqnarray}
K(x,y;t)=\left\{
\begin{array}{cl}
K_{++}(x,y;t)& x\in M_+,\; y\in M_+\\ K_{+-}(x,y;t)& x\in M_+,\;
y\in M_-\\K_{-+}(x,y;t) & x\in M_-,\; y\in M_+ \\ K_{--}(x,y;t) & x\in M_-,\;
y\in M_-
\end{array}
 \right.\qquad \quad
\label{eq:kkkk}
\end{eqnarray}
which satisfy the heat equations
\begin{eqnarray}
\left[\frac{\partial}{\partial
t}-a_{+}^2\;\Delta_x\right]\,\left\{K_{++} \atop
K_{+-}\right\}=0, \quad
\left[\frac{\partial}{\partial
t}-a_{-}^2\;\Delta_x\right]\,\left\{K_{-+} \atop
K_{--}\right\}=0
\label{eq3}
\end{eqnarray}
and are glued together by matching conditions
\begin{eqnarray}
K_{++}\biggl|_{\Sigma^+}=K_{-+}\biggr|_{\Sigma^-},
\quad \lambda_+\,\frac{\partial K_{++}}{\partial
n_x}\biggl|_{\Sigma^+}=\lambda_-\,\frac{\partial
K_{-+}}{\partial n_x}\biggl|_{\Sigma^-},\\
K_{+-}\biggl|_{\Sigma^+}=K_{--}
\biggr|_{\Sigma^-},\quad
\lambda_+\,\frac{\partial K_{+-}}{\partial
n_x}\biggl|_{\Sigma^+}=\lambda_-\,\frac{\partial
K_{--}}{\partial n_x}\biggl|_{\Sigma^-}.
\label{eq:match}
\end{eqnarray}
Choosing the suitable representation for the functions $K$~(\ref{eq:kkkk}) in terms of
simple and double layer potentials and making account for the matching conditions
(\ref{eq:match}) one obtains a system of integral equations
iteratively solvable with respect to  the densities  of simple  and
double $\mu$ and $\nu$ layers. The substitution of the densities
into the corresponding  the heat kernels gives
\begin{eqnarray}
K_{++}(x,y;t)&=&K^0_{++}(x,y;t)+2 a_+^2\sum\limits_{n=1}^{\infty}\int\limits_0^{t}
d\tau \int\limits_{\Sigma}dz
K^0_{++}(x,z;t-\tau)\,\mu_{n-1}(z,y;\tau),\nonumber\\
K_{-+}(x,y;t)&=&2 a_-^2\sum\limits_{n=1}^{\infty}\int\limits_0^{t}
d\tau \int\limits_{\Sigma}dz
\frac{\partial}{\partial n_z}K^0_{--}(x,z;t-\tau)\,\nu_{n-1}(z,y;\tau)
\label{eq:hkmre}
\end{eqnarray}
where
\begin{eqnarray}
\nu_i(x_{\Sigma},y;t)&=&2 a_+^2\int\limits_0^t d\tau\int\limits_{\Sigma}^{}dz\,
K^0_{++}( x_{\Sigma},z;t-\tau)\,\mu_{i-1}(z,y;\tau) \label{eq:nu}\\
&&-2 a_-^2\int\limits_0^t d\tau\int\limits_{\Sigma}^{}dz\,
\frac{\partial}{\partial n_z}K^0_{--}({\bf x},z;t-\tau)\,\nu_{i-1}(z,y;\tau),
\nonumber\\
\mu_i( x_{\Sigma},y;t)&=&-2 a_+^2\int\limits_0^t d\tau\int\limits_{\Sigma}^{}dz\,
\frac{\partial}{\partial n_x}\,K^0_{++}(x,z;t-\tau)\Biggl|_{x_{\Sigma}}\,
\mu_{i-1}(z,y;\tau) \label{eq:mu}\\&&+2\,a_-^2\frac{\lambda_-}{\lambda_+}
\int\limits_0^t d\tau \int\limits_{\Sigma}^{}dz\,
\frac{\partial}{\partial n_x}\,\frac{\partial}{\partial n_z}\,
K^0_{--}( x,z;t-\tau)\Biggl|_{x_{\Sigma}}\,\nu_{i-1}(z,y;\tau),\nonumber
\end{eqnarray}
\begin{eqnarray*}
\nu_0(x_{\Sigma},y;t)=K^0_{++}(x_{\Sigma},y;t), \quad
\mu_0(x_{\Sigma},y;t)=-\frac{\partial}{\partial n_x}\,K^0_{++}(x,y;t)
\Biggl|_{x_{\Sigma}}
\end{eqnarray*}
with $K^0_{\pm\pm}(x,y;t)=(4\pi a_{\pm}^2 t)^{-3/2}\exp\{(x-y)^2/(4 a_{\pm}^2 t)\}$.

The solution for the second pair of the heat kernels $K_{--}$ and $K_{+-}$
is obtained by replacing $K_{++}^{0}\leftrightarrow K_{--}^{0}$, and
$\lambda_+ \leftrightarrow \lambda_-$ in the righthand sides of (\ref{eq:hkmre}).
The functions $K_{++}$, $K_{-+}$, $K_{+-}$, and $K_{--}$ define the heat kernel
in the whole space. For details see~\cite{PBN}.

We are interested in the  asymptotic expansion of the heat kernel trace
$K(t)=\int_M dx \,K(x,x;t)$ when $t\to 0$. The functions $K_{-+}$ and $K_{+-}$ do
not contribute to it, thus we have to consider only $K_{++}$ and $K_{--}$.

For our purposes it is convenient to use such coordinates that
in the vicinity of the surface $\Sigma$ the metric is
$g_{ij}\,dx^i\,dx^j=(dx^3)^2+g_{ab}\,dx^a\,dx^b $
where $x^3$ is a coordinate on the normal to $\Sigma$, $x^3$=0 on
$\Sigma$.

Performing the surface integrations we keep in mind that for small $t$ the
contributions of largely separated points
are exponentially damped.  Therefore in the
vicinity of $\Sigma $ we may replace the squared distance $(x-z)^2$ by several
terms
of its expansion in powers of the geodesic distance $\sigma$ on the surface $\Sigma$
\begin{eqnarray}
(x-z)^2&=&(x_3-z_3)^2+\sigma^2 \left\{1-(x_3+y_3) k_1+
x_3\,z_3 (k_1^2+k_2^2)\right\} \nonumber\\
&&+\sigma^3\left\{-\frac{1}{3}(2 z_3+x_3)k_1'+
x_3 z_3 (k_1 k_1'+k_2 k_2') \right\}+\dots
\label{eq:dist}\\
k_1=L_{ab}\xi^a\xi^b, && k_2=\frac{1}{2}(e_{a\gamma}L^{\gamma}_b+
e_{b\gamma}L^{\gamma}_a)\,\xi^a \xi^b,
\quad k_1'\equiv\frac{d k_1}{d\sigma},\;\;k_2'\equiv\frac{d k_2}{d\sigma}.\nonumber
\end{eqnarray}

The surface area element is
$d z=\left(1-\frac{1}{12}r_{ab}\xi^a\xi^b\,\sigma^2+\dots\right)\sigma\,
d\sigma \,d\Omega,$
 $\Omega$ parameterises a unit sphere,
$L_{ab}$ is the second fundamental form on $\Sigma$, $r_{ab}$ is
intrinsic Ricci curvature, $\xi$ is a unit tangent vector at $x$ to the geodesics
with the length $\sigma$ joining $z$ to $x$ on $\Sigma$.

Here we display the results for the leading terms of the MRE when $t\to0$
\begin{eqnarray}
K^{(0)}(t)&=&K^{(0)}_{++}(t)+K^{(0)}_{--}(t)=\frac{t^{-3/2}}{(4\pi a_+^2)^{3/2}}\,M_+ +
\frac{t^{-3/2}}{(4\pi a_-^2)^{3/2}}\,M_-, \nonumber\\
K^{(1)}_{++}(t)&=&\frac{t^{-1}\,\Sigma}{8 \pi a_+^2}+
\frac{t^{-1/2}}{8\pi^{3/2}a_+}\int\limits_{\Sigma}L_{aa}+
\frac{t^0}{2^8\,\pi }\int\limits_{\Sigma}[5\,(L_{a}^{a})^2+L_{a}^{b}L_{b}^{a}-
\frac{2}{3}r^a_a]
+...,\nonumber\\
K^{(2)}_{++}(t)&=&-\frac{ t^{-1}}{8 \pi}\,\frac{\lambda_-}{\lambda_+}\,
\frac{\Sigma}{a_{+}^2}-\frac{1}{8 {\pi}^{3/2}}\frac{\lambda_-}{\lambda_+}
\frac{t^{-1/2}}{a_+}
\int\limits_{\Sigma}L_{aa}\nonumber\\
&&+\frac{t^0}{32
\,\pi}\Biggl\{
a_-\,
\frac{\lambda_-}{\lambda_+}\frac{(a_++2\,a_-)}{(a_++a_-)^2}
\int\limits_{\Sigma}[-(L_{a}^{a})^2+4\,L_{a}^{b}L_{b}^{a}-r_a^a/3] \nonumber\\
&&+
\Biggl[\frac{1}{8}-
\frac{\lambda_-}{\lambda_+}\frac{1}{(a_++a_-)^3}\biggl(\frac{35}{12}
a_-^3+
\frac{11}{4}\,a_-^2\,a_+ +2 a_+^2\,a_- \nonumber\\
&&+\frac{2}{3}\,a_+^3
+\frac{9}{4}\frac{a_-^4}{a_+}+\frac{3}{4}\frac{a_-^5}{a_+^2}\biggr)\Biggr]
\int\limits_{\Sigma}[(L_{a}^{a})^2+2\,L_{a}^{b}L_{b}^{a}]\Biggl\}+\dots
\end{eqnarray}
To obtain $K_{--}^{(1)}(t)$ and $K_{--}^{(2)}(t)$ one should replace $a_+ \leftrightarrow a_- $,
$\lambda_+\leftrightarrow\lambda_-$,
$L_a^b\rightarrow-L_a^b$. The asymptotic behaviour of the subsequent MRE terms may be
found in a similar way.  After that all factors appearing with the same powers of $t$
are added up to give the heat kernel coefficients. The latter are expressed though the
integrals of the surface geometric invariants.

\section{Conclusion}
In the present report we have shown how to formulate dielectric-like
spectral problem in terms of integral equations and to find the heat kernel
as their iterative solution. Our results indicate that this method
being a generalization of the known multiple reflection expansion~\cite{Moss,JH,Sant}
will finally help to derive the heat kernel coefficients  for the electromagnetic field
interacting with a dielectric body of arbitrary shape.

\section*{Acknowledgments}
This work was supported by RFBR grant N 03-01-00025, INTAS 2000-587. The author
is grateful to the organizers of  ${\cal Q}$FE{\large $\mathcal \chi$}$\mbox{T}_{03}$
and to UIPAP Women in Physics Travel Grant Program
for making her participation in the conference possible.

\end{document}